\documentclass[12pt,preprint]{aastex}

\shorttitle{SHARC-II observations of z$\geq$5 Quasars}
\shortauthors{Wang et al.}

\begin{document}


\title{SHARC-II 350$\mu$m Observations of Thermal Emission from Warm Dust in z$\geq$5 Quasars}


\author{Ran Wang\altaffilmark{1,2}, 
Jeff Wagg\altaffilmark{2}, 
Chris L. Carilli\altaffilmark{2}, 
Dominic J. Benford\altaffilmark{3}, 
C. Darren Dowell\altaffilmark{4}, 
Frank Bertoldi\altaffilmark{5}, 
Fabian Walter\altaffilmark{6}, 
Karl M. Menten\altaffilmark{7}, 
Alain Omont\altaffilmark{8}, 
Pierre Cox\altaffilmark{9}, 
Michael A. Strauss\altaffilmark{10}, 
Xiaohui Fan\altaffilmark{11}, 
Linhua Jiang\altaffilmark{11}}
\altaffiltext{1}{Department of Astronomy, Peking University, Beijing 100871, China}
\altaffiltext{2}{National Radio Astronomy Observatory, PO Box 0, Socorro, NM, USA 87801}
\altaffiltext{3}{NASA / Goddard Space Flight Center, Code 665 - Observational Cosmology Lab, Greenbelt, MD, USA 20771}
\altaffiltext{4}{Jet Propulsion Laboratory, Mail Code 169-506, Pasadena, CA, USA 91109}
\altaffiltext{5}{Argelander-Institut f$\rm \ddot u$r Astronomie, University of Bonn, Auf dem H$\rm \ddot u$gel 71, 53121 Bonn, Germany}
\altaffiltext{6}{Max-Planck-Institute for Astronomy, K$\rm \ddot o$nigsstuhl 17, 69117 Heidelberg, Germany} 
\altaffiltext{7}{Max-Plank-Institute for Radioastronomie, Auf dem H$\rm \ddot u$gel 71, 53121 Bonn, Germany} 
\altaffiltext{8}{Institut d'Astrophysique de Paris, CNRS and Universite Pierre et Marie Curie, Paris, France} 
\altaffiltext{9}{Institute de Radioastronomie Millimetrique, St. Martin d'Heres, F-38406, France}
\altaffiltext{10}{Department of Astrophysical Sciences, Princeton University, Princeton, NJ, USA, 08544} 
\altaffiltext{11}{Steward Observatory, The University of Arizona, Tucson, AZ 85721}

\begin{abstract}

We present observations of four $\rm z\ge5$ SDSS quasars 
at 350$\,\mu$m with the SHARC-II bolometer camera on the 
Caltech Submillimeter Observatory. These are among the 
deepest observations that have been made by SHARC-II at 350$\,\mu$m, 
and three quasars are 
detected at $\rm \ge3\sigma$ significance, greatly 
increasing the sample of 350$\,\mu$m (corresponds to rest 
frame wavelengths of $\rm <60\,\mu m$ at $\rm z\ge5$), 
detected high-redshift quasars. The derived rest frame far-infrared (FIR) 
emission in the three detected sources is about five to ten times 
stronger than that expected from the average SED of the local quasars 
given the same $\rm 1450\,\AA$ luminosity. Combining the previous 
submillimeter and millimeter observations at longer 
wavelengths, the temperatures of the FIR-emitting warm dust 
from the three quasar detections are estimated to be in 
the range of $\rm 39$ to $\rm 52\,K$. 
Additionally, the FIR-to-radio SEDs of the three 
350$\,\mu$m detections are consistent with the emission 
from typical star forming galaxies.
The FIR luminosities are $\rm \sim10^{13}\,L_{\odot}$ and 
the dust masses are $\rm \ge 10^{8}\,M_{\odot}$. These 
results confirm that huge amounts of warm dust can exist in 
the host galaxies of optically bright quasars as 
early as z$\rm \sim$6. The universe is so young  
at these epochs ($\rm \sim1\,Gyr$) that a rapid dust formation mechanism 
is required. We estimate the size of the FIR dust 
emission region to be about a few kpc, and further provide a comparison of the SEDs among  
different kinds of dust emitting sources to investigate the dominant 
dust heating mechanism.

\end{abstract}


\section{Introduction}

Quasars in the distant universe may be 
studied as important probes of supermassive black hole 
(SMBH) and host galaxy formation at early epochs. 
Universal relationships between the supermassive 
black holes (SMBHs) and their stellar bulges were 
found in both active and normal galaxies locally, 
indicating that the early evolution of the SMBH-bulge 
systems is tightly correlated (eg. Marconi \& Hunt 2002; 
Tremaine et al. 2002). How these relationships behaved 
at an early galaxy evolution stage is a critical  
question for the studies of the high-z universe. In particular, 
if the SMBH forms prior to the stellar bulge, we may expect 
to see massive star formation co-eval with rapid black hole 
accretion in the quasar systems at the 
highest redshifts. 

Observations at submillimeter and millimeter [(sub)mm] 
wavelengths were preformed to study the thermal emission from warm 
dust in the quasar host galaxies at high redshifts. 
These studies yield important information on the mass 
and temperature of the FIR-emitting dust, and thus provide key 
constraints on the related galaxy evolution activities. 
Samples of optically bright quasars from z$\sim$2 to 6 
have been observed at (sub)mm wavelengths (Omont et al. 1996; 
Omont et al. 2001; 2003, Carilli et al. 2001; Beelen et al. 2003; 
Bertoldi et al. 2003a; Priddey et al. 2003a, Robson et al. 2004; 
Wang et al. 2007), using the Max-Planck Millimeter 
Bolometer (MAMBO) on the IRAM-30m Telescope and the 
Submillimeter Common-User Bolometer Array (SCUBA) on the 
James Clerk Maxwell Telescope. The detection rate is 
about 30\% at mJy sensitivity for all these samples, 
indicating the existence of a population of FIR luminous 
objects within the high-z optically selected quasar sample. 

Important results were obtained from deep 350$\,\mu$m 
observations of these strong (sub)mm quasars (Benford et al. 1999; 
Beelen et al. 2006) made by the SHARC-II bolometer 
camera (Dowell et al. 2003) on the Caltech Submillimeter 
Observatory (CSO). The 350$\,\mu$m observations 
confirm the existence, and constrain the properties of the warm 
dust with a typical temperature of $\rm \sim50\,K$ in 
the host galaxies of these quasars (Benford et al. 1999;
Beelen et al. 2006). Fits of the FIR (SEDs) imply 
FIR luminosities of about $\rm \sim10^{13}\,L_{\odot}$ and 
FIR-to-radio luminosity ratios following the trend defined 
by star forming galaxies (Beelen et al. 2006). 
These properties are all comparable to the FIR emission
found in the submillimeter detected galaxies (SMGs)
at z$\sim$1 to $\gtrsim$3, i.e. thermal emission from $>$10 K to 70 K
warm dust heated by star formation at a rate of a few
$\rm 10^{3}\,M_{\odot}\,yr^{-1}$ (eg. Chapman et al. 2005; Kov$\rm \acute a$cs et al. 2006b).
The SHARC-II detected quasars also tend to have highly excited 
molecular CO emission from the host galaxies
(i.e. peak at $\rm J=6-5$ or higher, Carilli et al. 2002, 2007; 
Solomon \& Vanden Bout 2005), exhibiting FIR-to-CO 
emission ratios similar to that of the star forming galaxies 
at low and high redshifts (Solomon \& Vanden Bout 2005; 
Riechers et al. 2006; Carilli et al. 2007). 

One instrumental finding among these observations is the SHARC-II 
detection of SDSS J114816.64+525150.3 at z=6.42 
(hereafter J1148+5251, Beelen et al. 2006). This is the 
first 350$\,\mu$m dust continuum detection beyond $\rm z=5$. 
The $\rm 4.5\times10^{8}\,M_{\odot}$ of $\rm \sim55K$ warm dust 
discovered in this source implies a similar evolution stage as 
was found in other (sub)mm detected quasars at lower 
redshifts, but exist at an epoch when the age of the 
universe is only $\sim$870 Myr. Probing the warm dust 
emission at such an epoch is critical as it constrains 
the time scale of dust formation in addition to the questions 
of dust heating and black hole-bulge evolution.
The detection of J1148+5251 indicates a rapid 
dust formation in these earliest galaxies (forming $\rm \ge10^{8}\,M_{\odot}$ 
of dust within $\le$1 Gyr), which requires confirmation 
with more examples at similar redshifts.

In this paper, we report our new 350$\,\mu$m SHARC-II
observations of four z$\ge$5 SDSS quasars. 
These quasars are among the strongest MAMBO $\rm 1.2\,mm$ detections 
at z$\ge$5 (Carilli et al. 2001; Petric et al. 2003; Priddey et al. 2003b;
Bertoldi et al. 2003a; Wang et al. 2007). 
The observations presented in this work are among the most sensitive  
ones that have been made by SHARC-II at 350$\,\mu$m, and are of critical 
importance as they deliver a data point on the short wavelength 
side or near the peak of the redshifted FIR emission spectrum. 
Combining with measurements on the long wavelength side allows determination 
of the dust temperature and, thus, the mass and the FIR luminosity. 
We describe the SHARC-II observations 
in Section 2, list the results in Section 3, and present 
discussions in Section 4. We adopt a $\rm \Lambda$-model 
cosmology with $\rm H_{0}=71km\ s^{-1}\ Mpc^{-1}$, $\rm \Omega_{m}=0.27$ 
and $\rm \Omega_{\Lambda}=0.73$ (Spergel et al. 2007), throughout this paper.

\section{Observations}

The observations were carried out with SHARC-II on the 10.4 m telescope of the CSO during 
2007 January 13-18 UT. The SHARC-II 
camera is a $\rm 12\times 32$ pixel array with a beam 
size (FWHM) of $\rm 8.5''$ at 350$\,\mu$m. The field of view 
is $\rm 2.6'\times1.0'$. We observed the four target 
sources when the opacity at 225GHz ($\rm \tau_{225GHz}$) 
was $\le$ 0.06, i.e. during the best weather conditions on Mauna Kea. 
The CSO Dish Surface Optimization 
System (DSOS, Leong 2005) was used to correct the surface 
imperfections and gravitational deformations. Scans were 
performed using a Lissajous pattern with amplitudes of $\rm \pm45''$ 
and $\rm \pm12''$ in azimuth and elevation respectively, resulting in 
a uniform coverage of $\rm \sim65''\times34''$. We performed hourly pointing, focus, 
and flux calibration with strong sources. We used Uranus as a 
primary calibrator, and a number of sources, IRC10216, CRL618, CIT6, 
OH231.8+4.2, GL490, and Arp220 were used as secondary calibrators. The uncertainties 
of the final calibration were expected to be within 20\%, based on 
repeated observations of the calibrators and the flux uncertainties  
of the secondary calibrators. The $\rm 1\sigma$ pointing uncertainties, estimated 
from repeated pointing on calibrators, is about $\rm 2.3''$. 

During standard observing, the position of the secondary mirror is
updated frequently to compensate for sag as the telescope elevation
changes. Unfortunately, a telescope configuration problem on the last
three nights resulted in only infrequent updates.  The effects of
this problem are a drift in the focus and a small drift in pointing.  We
carefully studied the effects of the mirror position errors on the data
using the position encoders and images of the calibrators.  About half of
the data was dropped due to bad focus.  For the remaining data, we made
small ($\rm <3''$) adjustments to the pointing with an uncertainty of $\rm <1''$. 

We completed data reduction using the SHARC-II data reduction 
package CRUSH version 1.52 (Kov$\rm \grave a$cs 2006a). The total 
integration time for each source is about 6 to 8 hours 
after excluding the bad scans, and the rms noise level for the final 
maps is about $\rm 5$ to $\rm 6\,mJy\,beam^{-1}$. The final 
maps were smoothed to a FWHM of $\rm 12.4''$, for optimal 
signal-to-noise in the case of point-source detection, and the peak surface
brightness is adopted as the total flux density of the source. 

\section{Results}

Three out of the four z$\ge$5 quasars observed are detected 
at the $\rm \gtrsim3\sigma$ level with SHARC-II. We list 
basic information and previous (sub)mm and radio 
continuum results of the four targets in Table 1,  
and the results of the SHARC-II observations are presented 
in the first four columns of Table 2, including the source 
name, 350$\mu$m peak surface brightness, 
peak offset from the optical quasar, and the integration time.
The final SHARC-II maps of all the
four targets are presented in Figure 1. The $\rm 1\sigma$ 
position uncertainty on the map is   
$\rm \sigma_{pos}\sim\frac{FWHM}{SNR} \sim 4''$ for $\rm 3\sigma$ 
detections (considering the smoothed beam size of $\rm FWHM=12.4''$). 
Thus the 350$\,\mu$m peak positions of the three detections are 
consistent with the optical quasars given this position 
uncertainty (See Column 3 in Table 2).
The details for the individual sources are presented in the following.

{\bf J033829.31+002156.3} (hereafter J0338+0021) is a z=5.03 quasar discovered by Fan et al. (1999). It is one of 
the strongest MAMBO detections among a sample of z$\rm \ge4$ SDSS quasars 
from Carilli et al. (2001) with $\rm S_{\nu,1.2mm}=3.7\pm0.3\,mJy$. The dust 
continuum is also detected by SCUBA at 850$\,\mu$m (Priddey et al. 2003b). 
Moreover, Maiolino et al. (2007) detected CO(5-4) line emission in 
this source, with a molecular gas mass of $\rm 2.4\times10^{10}\,M_{\odot}$. 

This source is detected by our SHARC-II observation at 
the $\rm \sim4\sigma$ level, with a peak surface 
brightness of $\rm S_{\nu,350\mu m}=17.7\pm4.4\,mJy\,beam^{-1}$ $\rm 3.8''$ away from the optical position.

{\bf J075618.14+410408.6} (hereafter J0756+4104) was discovered by Anderson et al. (2001) with a redshift of z=5.09. 
It is a strong MAMBO source with $\rm S_{\nu,1.2mm}=5.5\pm0.5\,mJy$ 
(Petric et al. 2003), and also is detected at 450$\,\mu$m and 850$\,\mu$m by 
SCUBA (Priddey et al. 2003b; Priddey et al. 2007) and at 1.4GHz by the VLA (Petric et al. 2003). 

We detect a $\rm 3.3\sigma$ peak ($\rm S_{\nu,350\mu m}=17.1\pm5.2\,mJy\,beam^{-1}$) 
on the SHARC-II map, $4.3''$ away from the position of the optical quasar.

{\bf J092721.82+200123.7} (hereafter J0927+2001) is a SDSS quasar at z=5.77 
(Fan et al. 2006b; Carilli et al. 2007). This source has a  
MAMBO flux density of $\rm S_{\nu,1.2mm}=5.0\pm0.8\,mJy$, making it one of 
the most luminous quasars at z$\sim$6 at FIR wavelengths (Wang et al. 2007). 
The VLA radio continuum observation of this source 
yields a $\rm 3\sigma$ detection of $\rm S_{1.4GHz}=45\pm14\,\mu Jy$ at 1.4GHz. 
CO(5-4) and CO(6-5) line emission has also been detected in this source 
very recently (Carilli et al. 2007). The CO luminosity 
indicates molecular gas with a mass of $\rm 1.6\times10^{10}\,M_{\odot}$ 
in the host galaxy.
The CO observations are accompanied with a $\rm \sim3\sigma$
continuum measurement at $\rm 3.5\,mm$ of $\rm S_{\nu,3.5mm}=0.12\pm0.03\,mJy$.

With SHARC-II we measure a brightness  
for this source of $\rm S_{\nu,350\mu m}=17.7\pm5.7\,mJy\,beam^{-1}$ with a position 
offset of $\rm 3.9''$ from the optical quasar. A second $\rm \sim3\sigma$ peak is seen in 
the map of this source with $\rm S_{\nu,350\mu m}=17.5\pm5.6\,mJy\,beam^{-1}$, to 
the southeast of the optical quasar position. The separation between the two 
peaks is about $\rm 14.9''$. 
However, there is no hint of extension towards the secondary peak 
direction in the CO emission maps of J0927+2001 at $\rm 3\,mm$ of 
Carilli et al. (2007). Further high resolution mapping 
at (sub)mm wavelengths is required to understand the nature 
of this secondary peak. The analysis below only considers the first peak 
close to the optical quasar position. 

{\bf J104845.05+463718.3} (hereafter J1048+4637) is a broad absorption line quasar 
at z=6.20 from Fan et al. (2003). 
The MAMBO flux density for this source is $\rm S_{\nu,1.2mm}=3.0\pm0.4\,mJy$ 
(Bertoldi et al. 2003). However, it has not been detected by 
SCUBA (Robson et al. 2004), or by the VLA (Wang et al. 2007). 

Our SHARC-II observation did not detect this source. 
The pixel flux density value at the optical quasar position 
is $\rm 5.3\,mJy\,beam^{-1}$, with a $\rm 1\sigma$ rms 
of $\rm 5.8\,mJy\,beam^{-1}$ on the final map. Thus we adopt 
$\rm 17.4\,mJy\,beam^{-1}$ ($\rm 3\sigma$ rms) as 
an upper limit to the 350$\,\mu$m flux density. 

The optical-to-radio SEDs of 
the four quasars are plotted in Figure 2, displaying the 
photometric data listed in Table 1 and 2. Template SEDs of 
local quasars from Elvis et al. (1994) and 
Richards et al. (2006) are plotted for comparison, 
normalized to $\rm 1450\AA$. 
For J1048+4637, the measurements at submillimeter, 
millimeter and radio wavelengths are all consistent with the 
templates, i.e. the average SED emission of the typical 
type-I quasar in the local universe. However, strong 
FIR emission is seen in the other three sources, which 
exceed the templates by nearly an order of magnitude. 
This is similar to the FIR excess from warm dust shown 
in the SED of J1148+5251 by Beelen et al. (2006). 

We fit the FIR bumps of the three sources with an 
optically thin graybody, namely\\
\begin{equation}
\rm S_{\nu}=S_{0}\cdot\left(\frac{\nu}{1GHz}\right)^{3+\beta}\frac{1}{exp(h\nu/kT_{d})-1},\\
\end{equation}
where $\rm \beta$ is the emissivity index, $\rm T_d$ is the 
dust temperature in K, $\rm S_{0}$ is the amplitude factor, 
and $\rm S_{\nu}$ is the rest frame flux density in mJy.
Since there are only two or three data points 
available for each source, we fix the $\rm \beta$ value 
to 1.6 (Beelen et al. 2006), and then fit $\rm S_{0}$ 
and $\rm T_d$ for the three sources. A Levenberg-Marquardt 
least-squares fit is performed with the IDL package MPFIT. 
The best-fitting dust temperatures range from $\rm 39\ to\ 52\,K$ 
with $\rm 1\sigma$ fitting errors of 3 to 4 K. In Figure 2, we 
also extended the fitted FIR SED to radio for the three detections, 
using a radio spectral index of -0.75 and an average  
FIR-to-radio emission ratio of $\rm q\equiv log(L_{FIR}/3.75
\times10^{12}\,W)-log(L_{\nu,1.4GHz}/W\,Hz^{-1})=2.34$ found 
in typical star forming galaxies (Yun et al. 2001). 
The observed radio emission/upper limit of the three objects 
are all consistent with the model SED, i.e. within the range of  
FIR-to-radio emission ratios five times 
above or below the average value of star forming galaxies, but the detected  
flux densities in J0756+4104 and J0927+2001 are 0.2 to 0.4 index higher 
than the value derived with q=2.34.

The FIR luminosity is calculated from the fitted 
SED, integrating rest frame 42.5$\,\mu$m to 122.5$\,\mu$m, namely\\
\begin{equation}
\rm L_{FIR}=2.49\times10^{-11}\cdot4\pi D_{L}^{2}\int S_{\nu}d\nu\\
\end{equation}
where $\rm D_{L}$ is the luminosity distance in Mpc and $\rm L_{FIR}$ is in $\rm L_{\odot}$.  
We calculate the dust masses following \\
\begin{equation}
\rm M_{dust}=\frac{L_{FIR}}{4\pi\int\kappa_{\nu}B_{\nu}d\nu}=8.33\times10^{14}\cdot\frac{S_{0}D_{L}^{2}}{\kappa_{0}}\ M_{\odot}\\
\end{equation} 
where $\rm B_{\nu}$ is the planck function 
and $\rm \kappa_{\nu}=\kappa_{0}(\nu/\nu_{0})^{\beta}$ is the dust 
absorption coefficient. We adopt $\rm \kappa_{0}=18.75\,cm^2g^{-1}$ 
at 125$\,\mu$m (Hildebrand 1983). 
The derived parameters for the three SHARC-II detected sources 
are listed in the last four columns in Table 2.

\section{Discussion}

We have detected three high redshift quasars with  
SHARC-II at 350$\,\mu$m. These are among 
the most sensitive observations made by SHARC-II, and increase 
the number of 350$\,\mu$m detected optical quasars 
at z$\ge$5 to four. These quasars are all strong 
detections in previous MAMBO and SCUBA observations 
at longer wavelengths (Carilli et al. 2001; 
Petric et al. 2003; Robson et al. 2004; Wang et al. 2007), 
and the measurements given 
by SHARC-II are close to the FIR emission peak 
from warm dust. By sampling over the dust peak, we 
provide reliable measurements of temperature, mass, 
and FIR luminosity of the FIR-emitting warm dust.
As with J1148+5251, (Bertoldi et al. 2003; Beelen et al. 2006), 
the three newly detected sources show that excess emission from huge amounts of warm 
dust can exist in some optically 
bright quasars at extreme redshifts. 

The derived dust masses are all $\rm \ge10^{8}\,M_{\odot}$. 
This result confirms the high heavy element abundance and rapid dust 
formation in massive quasar hosts at z$\ge$5 
(Priddey et al. 2003b; Bertoldi et al. 2003a) when 
the age of the universe was $\rm \lesssim1.2\,Gyr$. At this epoch, 
the standard mechanism of ISM dust formation, i.e. stellar winds 
from evolved low mass stars, is 
inefficient as it requires very long time scales ($\rm >1\,Gyr$). 
A process associated with the evolution of massive stars, 
(with a much shorter time scale), is likely required 
(eg. Morgan \& Edmunds 2003; Schneider et al. 2004; Maiolino et al. 2004).

All of the SHARC-II detected z$\geq$5 quasars exhibit FIR luminosities 
of $\rm \sim10^{13}\,L_{\odot}$. Under the optically
thin assumption (i.e. the optical depth $\rm \tau_{\nu}\ll1$ at FIR wavelengths), 
the size of the dust emission region can be roughly estimated 
as \\
\begin{equation}
\rm R_{dust}\sim[\frac{L_{FIR}}{(4\pi)(\pi\int\tau_{\nu}B_{v}(T_{d}\sim50K)\,d\nu)}]^{0.5}  
\end{equation}
where the emissivity index $\rm \beta=1.6$, $\rm \tau_{\nu}=(\nu/\nu_{c})^{\beta}$, and $\rm \nu_{c}$ is 
the critical frequency with $\rm \tau_{\nu_{c}}=1$. To satisfy the optically 
thin assumption, we adopt $\rm \nu_{c}=30\,THz$ (i.e. $\rm \lambda_{c}=10\,\mu m$). 
The derived $\rm R_{dust}$ is $\rm \sim5\,kpc$, which suggests the 
FIR-emitting warm dust emission region is on kpc scales in 
the quasar host. This is consistent with the scales of CO 
emission and [C II] 158$\,\mu$m emission found in J1148+5251 (Walter, et al. 2004, 2007, in prep.).

The dust temperatures derived from these z$\geq$5 quasars 
are in the range from $\rm 39$ to $\rm55\,K$, which 
is comparable to the values found in the bright (sub)mm  
quasars at lower redshifts (Benford et al. 1999; 
Beelen et al. 2006). They are also within the range of 
about 11 to 72 K found in samples of submillimeter 
selected galaxies at z$\sim$1 to 3.5, but a little higher 
than the mean value of $\sim$35 K (Chapman et al. 2005; Kov$\rm \acute a$cs et al. 2006b).
We combine the infrared to radio data of all four SHARC-II 
detected z$\geq$5 quasars and plot them in Figure 3. 
The combined SED is compared to that of another two 
sources, the z=3.9 quasar APM 08279+5255 (Irwin et al. 1998; 
Lewis et al. 1998, 2002; Downes et al. 1999; Egami et al. 2000; 
Beelen et al. 2006; Wagg et al. 2005; Wei$\rm \beta$ et al. 2007) and the 
local starburst galaxy M82 (Telesco \& Harper 1980; 
Klein et al. 1988; Hughes et al. 1990; Smith et al. 1990; 
Kr$\rm \ddot u$gel et al. 1990). The FIR emission from 
APM 08279+5255 is suggested to be dominated by a $\sim$200 K 
dust component on scales of a few hundred pc, 
i.e. a typical AGN heated hot dust torus 
(Wei$\rm \beta$ et al. 2007), while the FIR emission 
in the starburst galaxy M82 is clearly from starburst 
heated warm dust with a temperature of $\sim$45 K 
(Klein et al. 1998). The combined FIR-to-radio emission of the 
four quasars can be probed by the SED of M82 very well, but 
is quite different from the dusty AGN emission from 
APM 08279+5255, which peaks at a shorter wavelength ($\rm \sim20\,\mu m$)
and shows a much steeper infrared slope. Among the four z$\geq$5 
SHARC-II detections, only J1148+5251 have Spitzer measurements 
at near-IR wavelengths (Charmandaris et al. 2004; Jiang et al. 2006), and the flux 
densities are all much lower compared to the near-IR 
emission of APM 08279+5255.

The estimate of $\rm R_{dust}$ and the comparison of the 
IR SEDs may suggest a starburst origin of the 
strong FIR emission in these z$\ge$5 quasars detected by SHARC-II and 
other (sub)mm detectors. If this is the case, the star 
formation rates derived from the
FIR luminosities are all $\rm \gtrsim10^{3}\,M_{\odot}yr^{-1}$
(as listed in Col. (8), Table 2)\footnote{We estimate the star formation rate with
the empirical relationship from Kennicutt
(1998) assuming a standard initial mass function, i.e. $\rm SFR\sim 4.5L_{IR}\,M_{\odot}yr^{-1}$,
where $\rm L_{IR}$ is the infrared 
luminosity (8-1000$\,\mu$m) in unit 
of $\rm 10^{44}\,erg\,s^{-1}$, and 
is $\rm\sim 1.5L_{FIR}$ for warm dust emission.}, indicating 
an active bulge building in these quasar hosts.
We expect future observations with 
Spitzer and the Herschel Space Observatory to 
better constrain the dust emission SEDs of these 
objects, and high resolution imaging 
($\rm \sim 1''$) with the Atacama Large Millimeter
Array (ALMA) to constrain the size of the dust emission region
These future observations will give a better understanding 
of star formation and warm dust heating in these high-z 
bright (sub)mm quasars.

\acknowledgments

We thank Jonathan Bird at the CSO for preparing the opacity data. 
We thank Colin Borys for helpful advice on SHARC-II data reduction.
The CSO is supported by the NSF under AST-0540882. 
We acknowledge support from the Max-Planck Society
and the Alexander von Humboldt Foundation through the
Max-Planck-Forschungspreis 2005. The National Radio 
Astronomy Observatory is a facility of the National 
Science Foundation operated under cooperative agreement 
by Associated Universities, Inc.

{\it Facilities:} \facility{CSO (SHARC-II)}, \facility{IRAM (MAMBO)}, \facility{VLA}, \facility{SDSS}.

\clearpage

\clearpage
\begin{table}
{\scriptsize
\begin{center}
\caption{Summary of the previous observations \label{tbl-1}}
\begin{tabular}{cccccccl}
\hline \noalign{\smallskip}
SDSS name&z&$m_{1450\AA}$ &$\rm S_{\nu,450\mu m}$&$\rm S_{\nu,850\mu m}$&$\rm S_{\nu,1.2mm}$& $S_{\nu,3.5mm}$ &$\rm S_{1.4GHz}$\\
         & & & mJy & mJy & mJy& mJy & $\rm \mu$Jy\\
(1)&(2)&(3)&(4)&(5)&(6)&(7)&(8) \\
\noalign{\smallskip} \hline \noalign{\smallskip}
J033829.31+002156.3&5.03&20.01$^{a}$&5$\pm$16$^{b}$&11.9$\pm$2.0$^{b}$&3.7$\pm$0.3$^{c}$& ---&37$\pm$25$^{c}$\\
J075618.14+410408.6&5.09&20.15$^{d}$&16$\pm$5$^{l}$&13.4$\pm$1.0$^{l}$&5.5$\pm$0.5$^{e}$&---&65$\pm$17$^{e}$\\
J092721.82+200123.7&5.77&19.87$^{f}$&---&---&5.0$\pm$0.8$^{g}$&0.12$\pm$0.03$^{k}$&45$\pm$14$^{g}$\\
J104845.05+463718.3&6.20&19.25$^{h}$&7.6$\pm$11.7$^{i}$&2.3$\pm$2.2$^{i}$&3.0$\pm$0.4$^{j}$&---&6$\pm$11$^{g}$\\
\noalign{\smallskip} \hline
\end{tabular}\\
\end{center}
$^{a}$Fan et al. (1999). $^{b}$Priddey et al. (2003b). 
$^{c}$Carilli et al. (2001). $^{d}$Anderson et al. (2001). $^{e}$Petric et al. (2004). 
$^{f}$Fan et al. (2006). $^{g}$Wang et al. (2007). $^{h}$Fan et al. (2003). 
$^{i}$Robson et al. (2004). $^{j}$Bertodi et al. (2003a). $^{k}$Carilli et al. (2007). 
$^{l}$Priddey et al. (2007)
}\end{table}

\begin{table}
{\scriptsize
\begin{center}
\caption{Results and derived parameters \label{tbl-2}}
\begin{tabular}{cccccccc}
\hline \noalign{\smallskip}
SDSS name& $\rm S_{\nu,350\mu m}^{a}$ & Offsets &Int. time & $\rm L_{FIR}$ & $\rm M_{dust}$ & $\rm T_{dust}$ & SFR\\
         &$\rm mJy\,beam^{-1}$& $''$&hours & $\rm 10^{13}\,L_{\odot}$& $\rm 10^{8}\,M_{\odot}$&K & $\rm 10^{3}\,M_{\odot}yr^{-1}$\\
(1)&(2)&(3)&(4)&(5)&(6) & (7) & (8)\\
\noalign{\smallskip} \hline \noalign{\smallskip}
J033829.31+002156.3&17.7$\pm$4.4&3.8& 7.2 &0.92$\pm$0.17 &6.1 &45.6$\pm$3.2& 2.2  \\
J075618.14+410408.6&17.1$\pm$5.2&4.3& 6.3 &0.84$\pm$0.17 &12.1 &39.2$\pm$2.6& 1.9 \\
J092721.82+200123.7&17.7$\pm$5.7&3.9& 7.3 &1.21$\pm$0.28 &4.6 &51.1$\pm$4.2& 3.2 \\
J104845.05+463718.3& $\rm 5.3\pm5.8$&---& 8.0 &--- &--- &---  &--- \\
\noalign{\smallskip} \hline
\end{tabular}\\
\end{center}
$^{a}$ Note that the absolute calibration uncertainty of 20\% is not included in the rms.\\ 
We adopt an emissivity index of $\rm \beta=1.6$ here for all the calculations (Beelen et al. 2006).
}\end{table}

\clearpage

\begin{figure}
\epsscale{1.2}
\plottwo{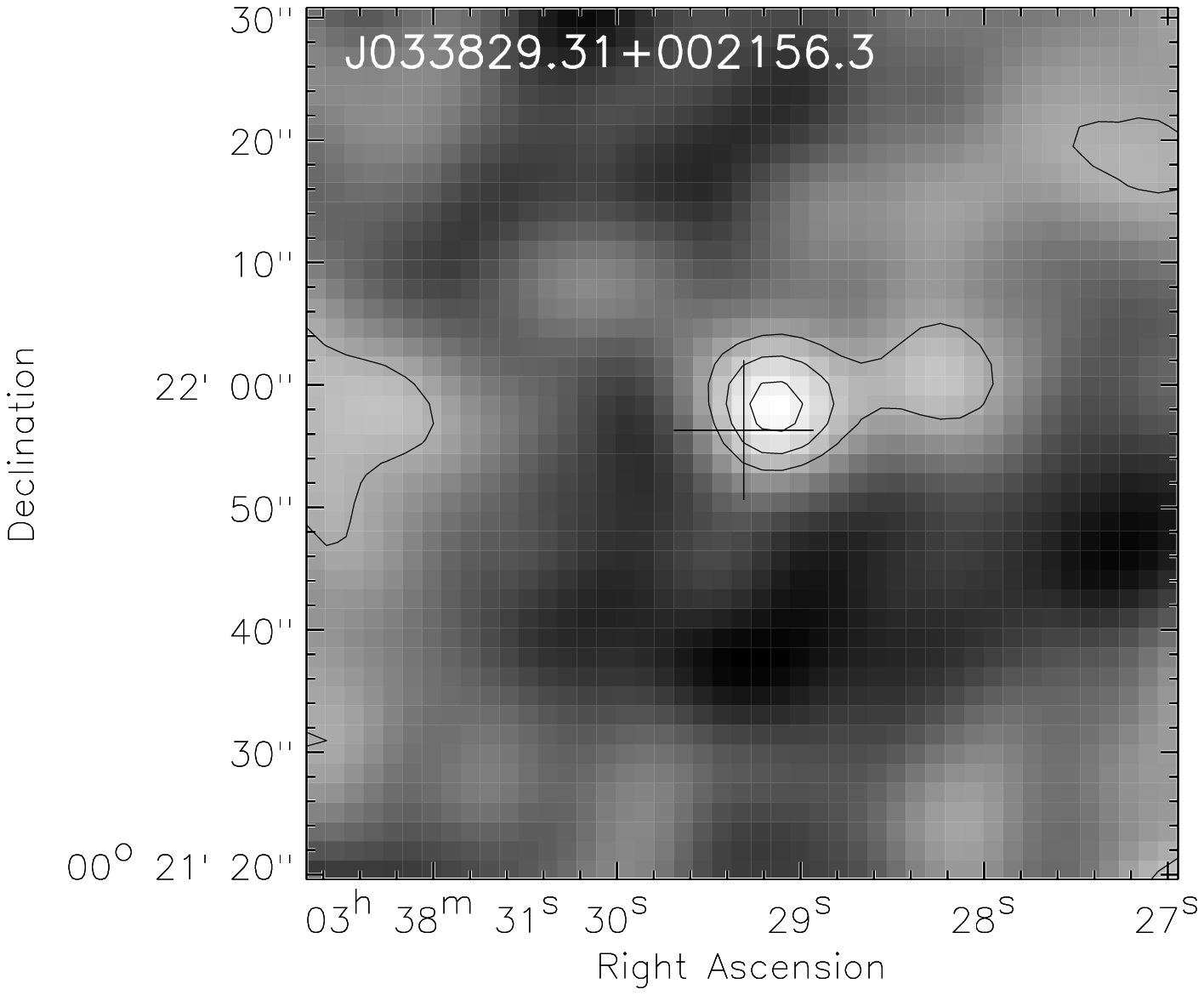}{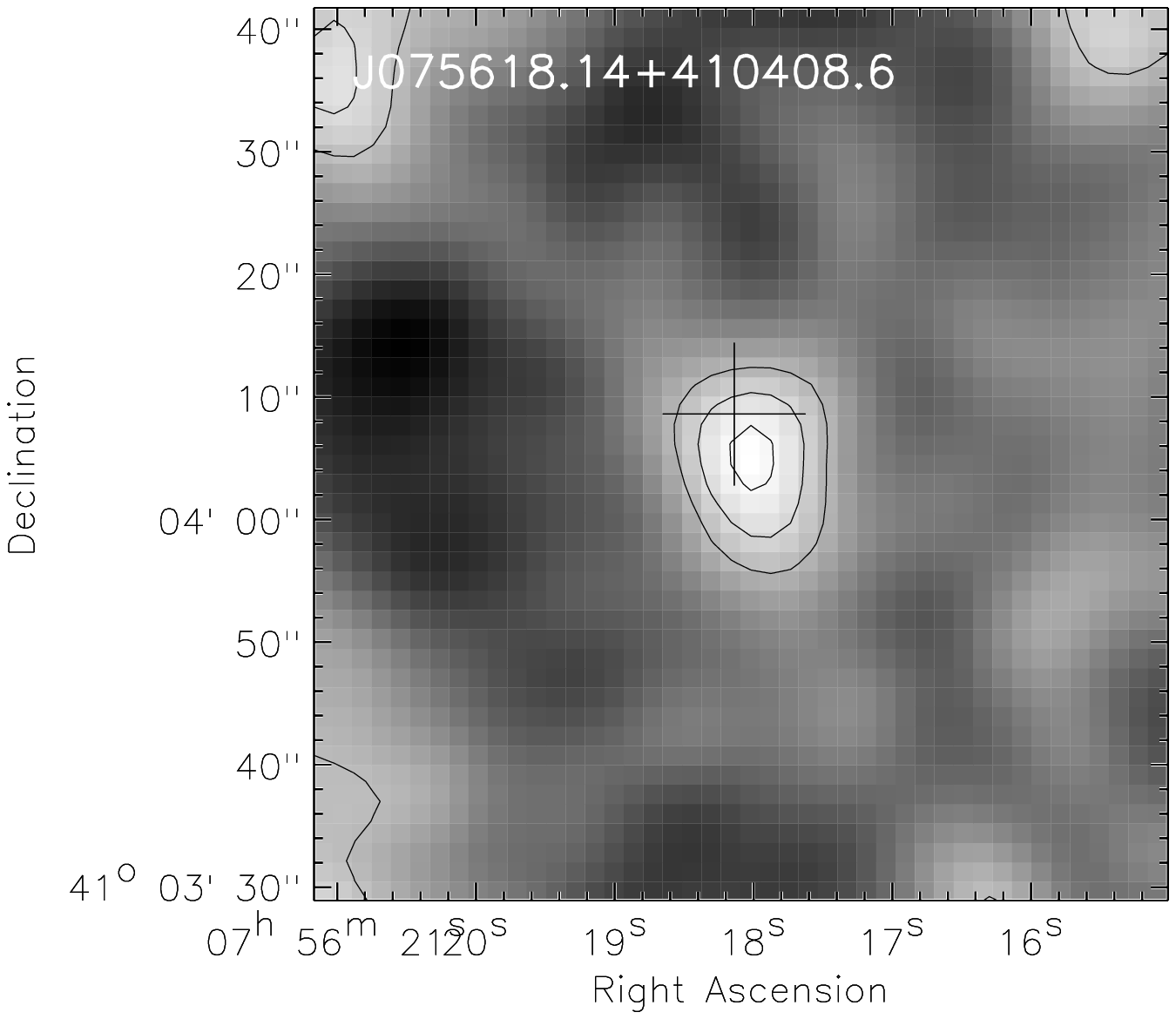}\\
\plottwo{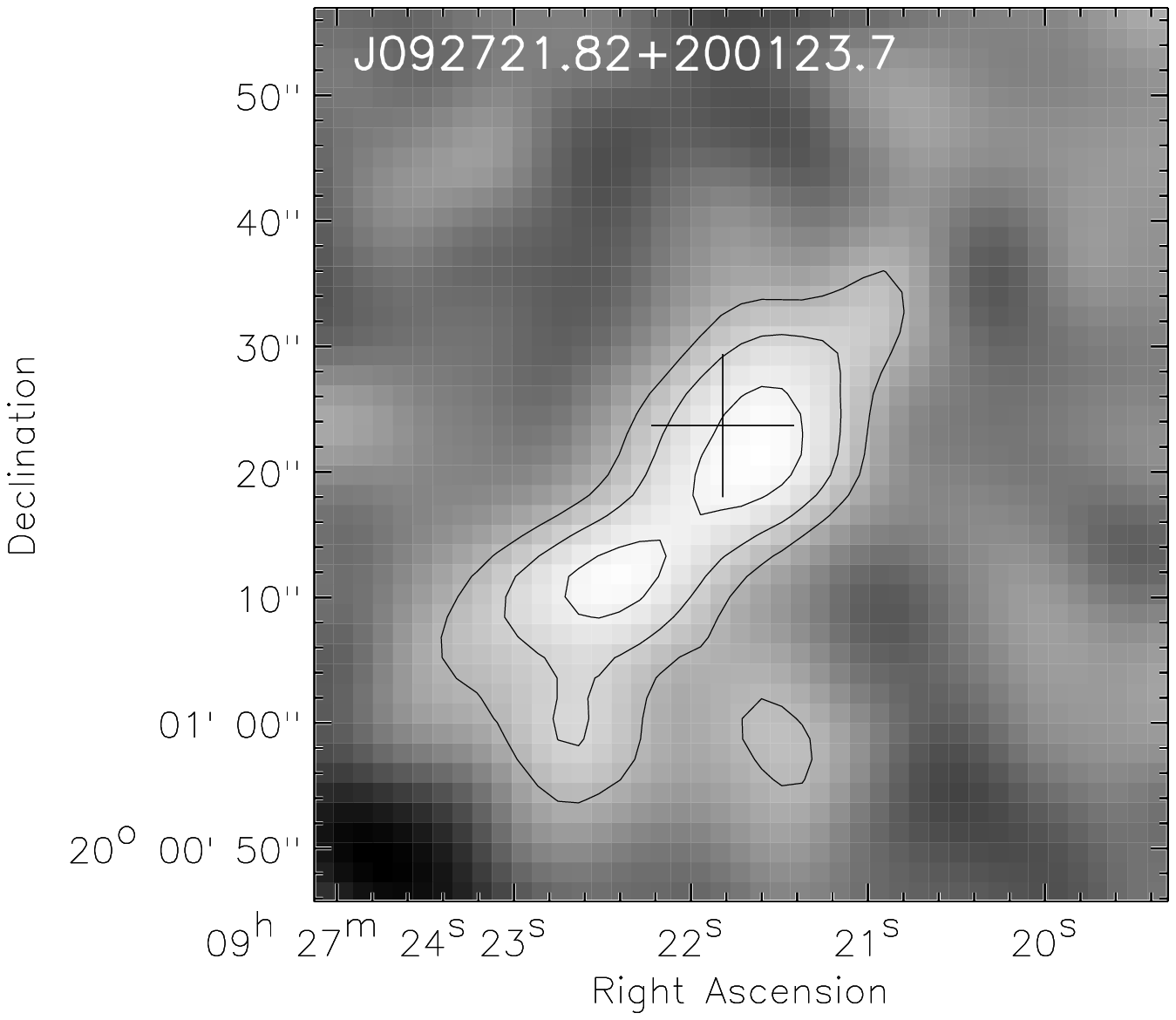}{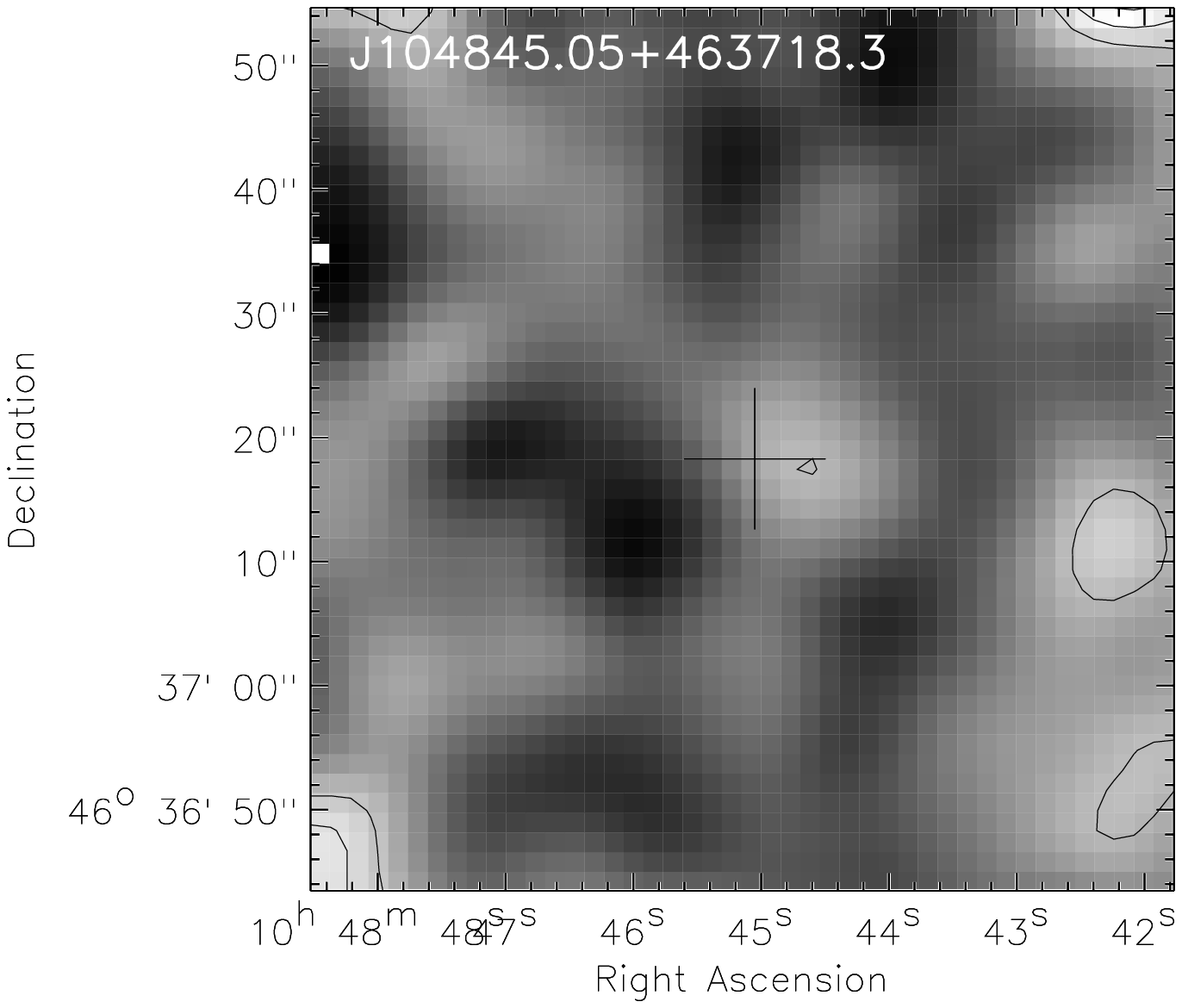}\\
\figcaption{The SHARC-II maps of the 
four z$\ge$5 quasars at 350$\,\mu$m, smoothing to a beam size 
of $\rm FWHM=12.4''$. The contour levels are 
$\rm (2,3,4)\times4.0\,mJy\,beam^{-1}$ for 
each of the four maps. The crosses mark the positions 
of the optical quasars. \label{fig1}}
\end{figure}

\begin{figure}
\epsscale{1.2}
\plottwo{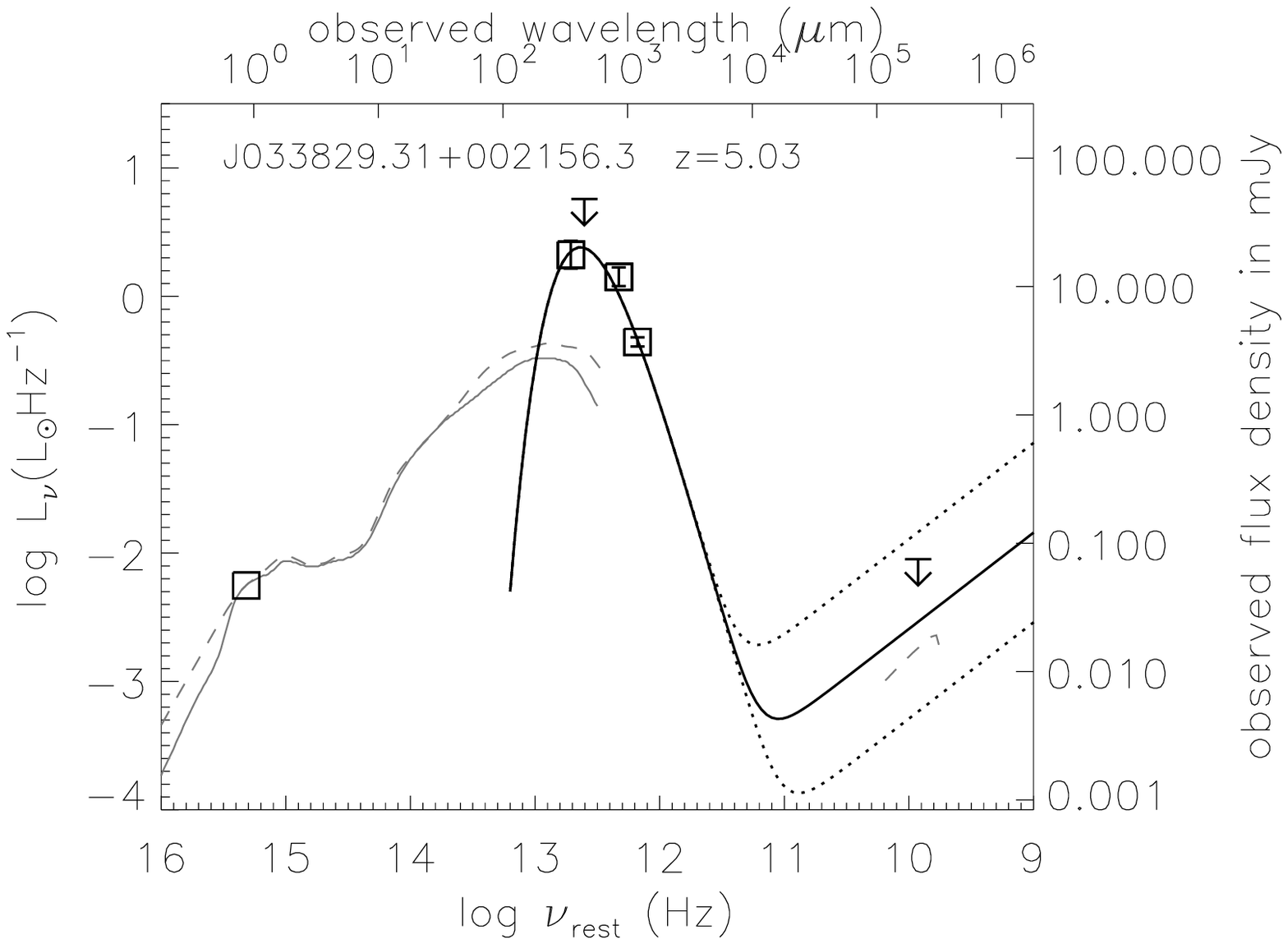}{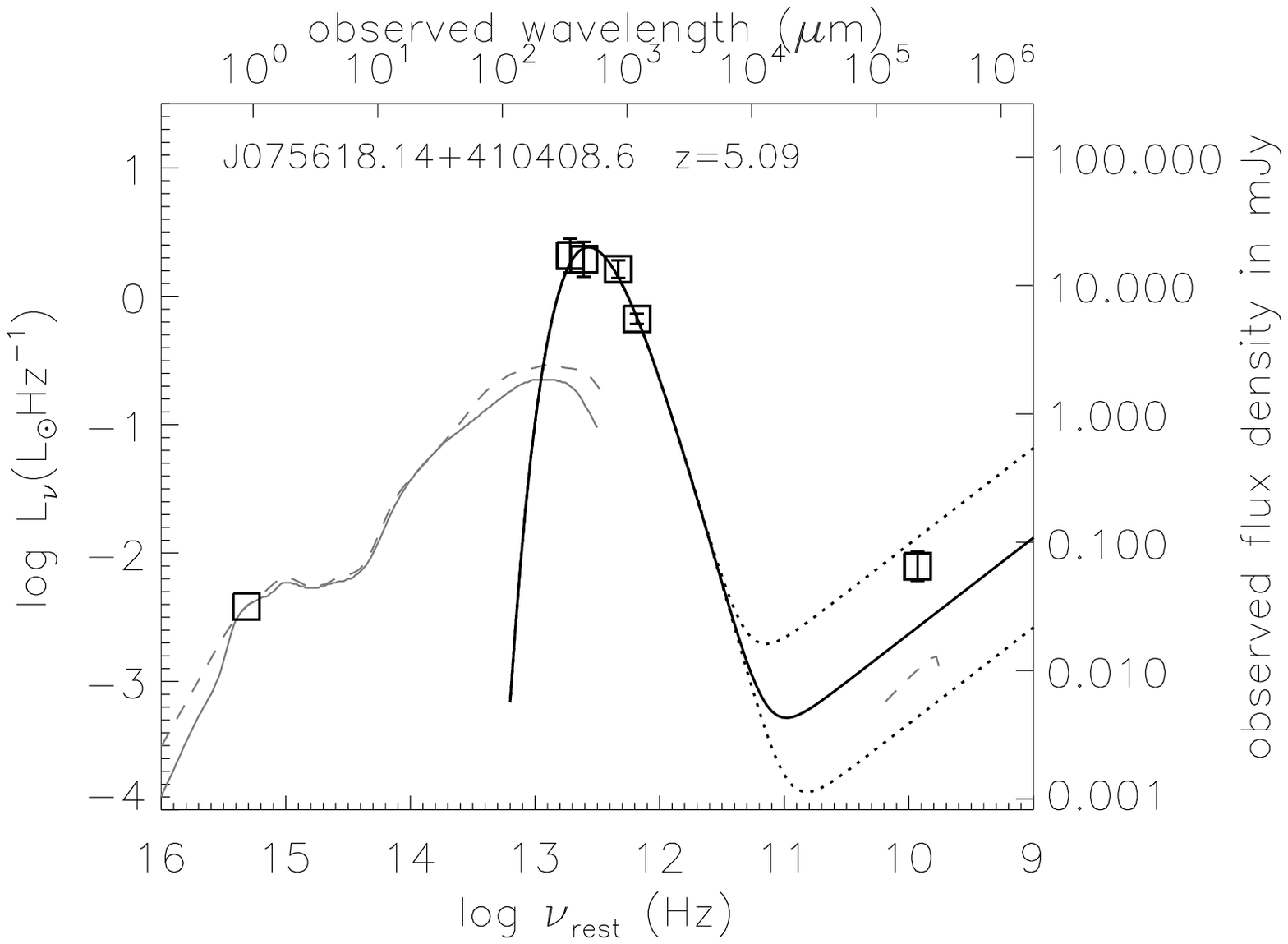}\\
\plottwo{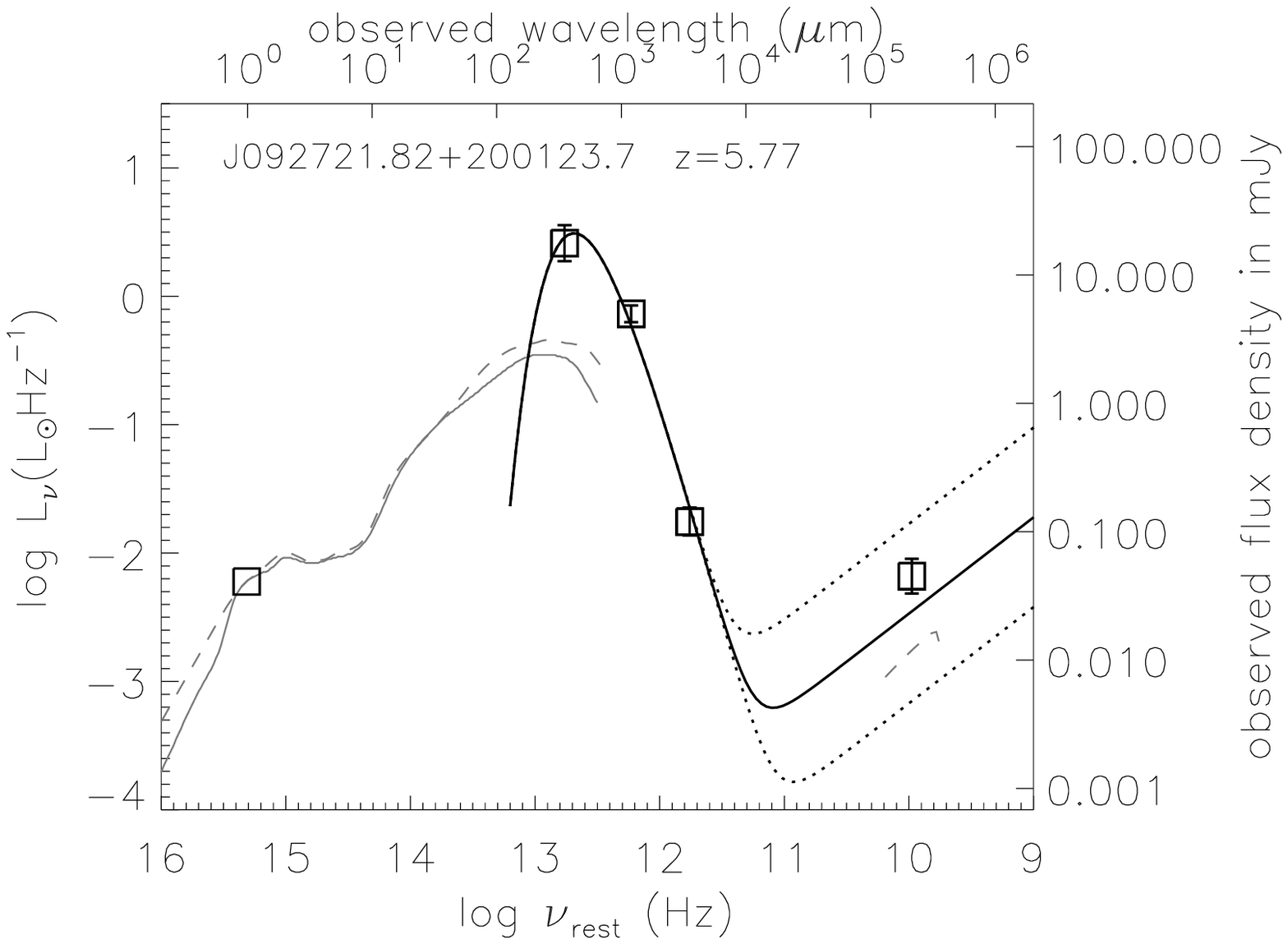}{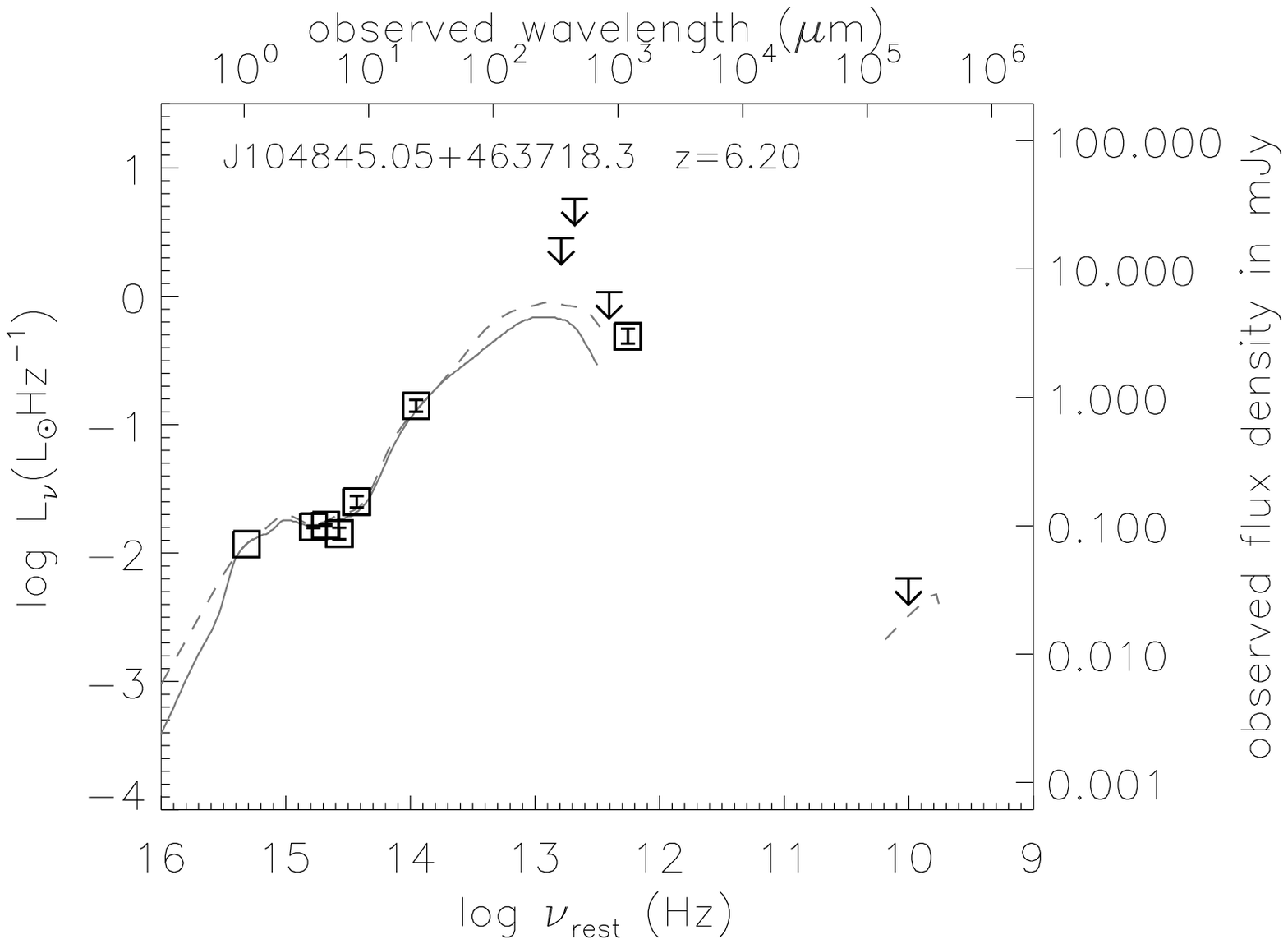}\\
\figcaption{Optical to radio SEDs for the four z$\ge$5 quasars. 
The squares represent optical data at 1450$\rm \AA$,
SHARCII data at 350$\,\mu$m, SCUBA data at 450$\,\mu$m 
and 850$\,\mu$m, MAMBO data at $\rm 1.2\,mm$, PdBI data at $\rm 3.5\,mm$, and VLA data at 1.4GHz.
Arrows denote $\rm 3\sigma$ upper limits for non-detections.
The dashed and solid lines represent the low-z quasar templates 
from Elvis et al. (1994) and Richards et al. (2006), respectively.
All the templates are normalized to 1450$\rm \AA$. The hard solid
line is the best-fit warm dust model for the three SHARC-II 
detections, extended to the radio band with the typical 
radio-FIR correlation of star forming galaxies, i.e. 
q=2.34 (Yun et al. 2001). The dotted lines denote factors 
of 5 excesses above and below this typical q value defined by star forming galaxies.}
\end{figure}

\begin{figure}
\epsscale{0.8}
\plotone{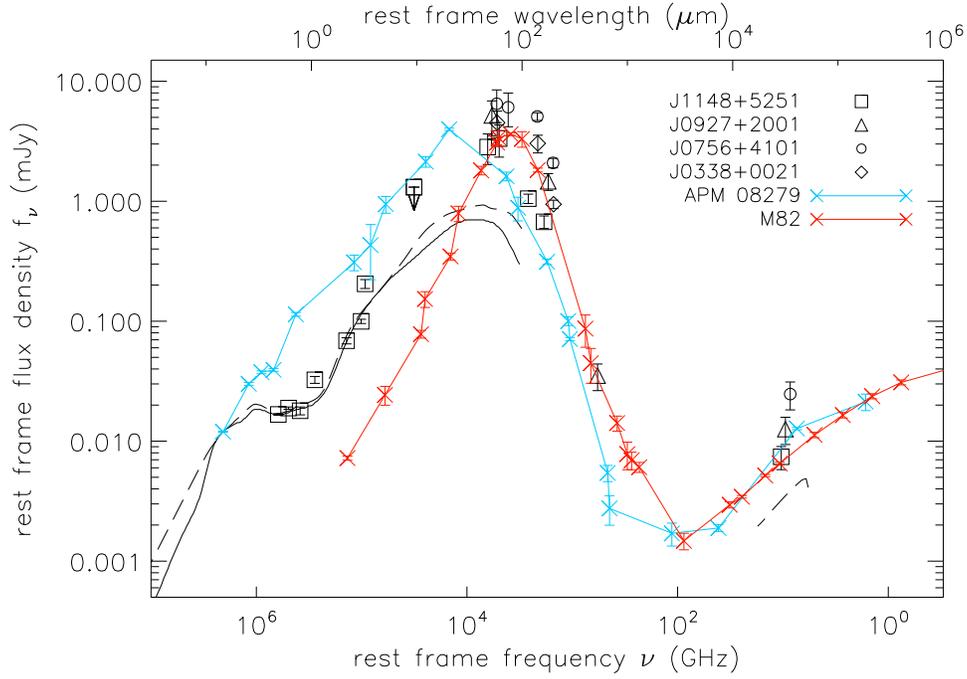}\\
\figcaption{Combined rest frame optical to radio SED of the four SHARC-II 
detected z$\ge$5 quasars. The data of J1148+5251 are taken from 
Jiang et al. (2006), Charmandaris et al. (2004), Beelen et al. (2006), Robson et al. (2004), 
Bertoldi et al. (2003a), and Carilli et al. (2004). 
The flux densities are normalized to 
the 1450$\rm \AA$ emission of J1148+5251. The black lines are quasar 
templates as described in Figure 2. The blue line is the SED of 
the $\rm z=3.9$ quasar APM 08279+5255, scaling to the 1450$\rm \AA$ 
emission of J1148+5251, with data from Irwin et al. (1998), 
Lewis et al. (1998, 2002), Downes et al. (1999), Egami et al. (2000), 
Beelen et al. (2006), Wagg et al. (2005) and Wei$\rm \beta$ et al. (2007).  
The red line is the SED of the central region of the starburst galaxy M82, scaling to the 0.5 mm 
flux density of the combined SED, with data from Telesco \& Harper (1980), 
Klein et al. (1988), Hughes et al. (1990), Smith et al. (1990), 
and Kr$\rm \ddot u$gel et al. (1990). }
\end{figure}




\end{document}